\PassOptionsToPackage{dvipsnames,svgnames,x11names}{xcolor}
\documentclass{trbunofficial}

\usepackage{amsfonts}
\usepackage{array}
\usepackage{booktabs}
\usepackage{multirow}
\usepackage{placeins}
\usepackage{xcolor}
\usepackage{tikz}
\usepackage{soul}
\usepackage[
  authordate,
  backend=biber,
  sorting=nyt,
  ibidtracker=false,
  maxcitenames=2,
  maxbibnames=99,
  doi=true,
  isbn=false,
  url=true,
  dashed=false
]{biblatex-chicago}
\AtBeginDocument{\let\cite\parencite}

\usetikzlibrary{positioning,arrows.meta,shadows.blur,backgrounds,fit,calc}

\definecolor{setupblue}{HTML}{DCE9FA}
\definecolor{processorange}{HTML}{FDE6CC}
\definecolor{resultgreen}{HTML}{DFF3DE}
\definecolor{darkblue}{HTML}{1F4E8C}
\definecolor{excludered}{HTML}{B03A2E}

\begin{document}

\title{Deployment Feasibility Analysis of Post-Quantum Digital Signatures in Safety-Critical C-V2X Communication for Urban Mobility Scenario}

\TRBauthor*{Akid Abrar}{Ph.D. Student\\Department of Civil, Construction, and Environmental Engineering\\The University of Alabama}{aabrar@crimson.ua.edu}[Tuscaloosa, AL 35487-0288]
\TRBauthor{Sagar Dasgupta}{Research Engineer\\Department of Civil, Construction, and Environmental Engineering\\The University of Alabama}{sdasgupta@ua.edu}[Tuscaloosa, AL 35487-0288]
\TRBauthor{Abdullah Al Mamun}{Ph.D. Student\\Glenn Department of Civil Engineering\\Clemson University}{abdullm@clemson.edu}[Clemson, SC 29634]
\TRBauthor{Minhaj Uddin Ahmad}{Department of Civil, Construction and Environmental Engineering, The University of Alabama}{mahmad12@crimson.ua.edu}[Tuscaloosa, AL, 35487-0288]
\TRBauthor{Mizanur Rahman}{Assistant Professor\\Department of Civil, Construction, and Environmental Engineering\\The University of Alabama}{mizan.rahman@ua.edu}[Tuscaloosa, AL 35487-0288]
\TRBauthor{Mashrur Chowdhury}{Eugene Douglas Mays Chair of Transportation\\Glenn Department of Civil Engineering\\Clemson University}{mac@clemson.edu}[Clemson, SC 29634]
\TRBauthor{Ahmad Alsharif}{Assistant Professor\\Department of Computer Science\\The University of Alabama}{aalsharif1@ua.edu}[Tuscaloosa, AL 35487-0290]

\AuthorHeaders{Abrar, Dasgupta, Mamun, Ahmad, Rahman, Chowdhury, and Alsharif}
\renewcommand{\trbdate}{July 31, 2026}
\maketitle

\section{Abstract}
\noindent\textbf{Objectives:} The transition from the classical Elliptic Curve Digital Signature Algorithm (ECDSA) to post-quantum cryptography (PQC) creates substantially larger authentication payloads for safety-critical cellular vehicle-to-everything (C-V2X) sidelink communication. This study determines which National Institute of Standards and Technology (NIST) post-quantum signature algorithms are compatible with the current SAE J3161 deployment profile and quantifies their communication-level effects.

\hfill\break%
\noindent\textbf{Methods:} A transport-block feasibility analysis was performed using IEEE 1609.2 secured-message structures, SAE J3161 radio parameters, and the signature and public-key sizes of ECDSA P-256, Falcon-512, Dilithium-2, and SPHINCS+. Falcon-512, the only post-quantum candidate that fit the applicable transport-block constraints, was compared with ECDSA P-256 through full-stack C-V2X PC5 Mode 4 co-simulation. The evaluation covered 24 scenarios spanning six traffic levels-of-service with line-of-sight and non-line-of-sight propagation. Packet delivery ratio (PDR) and end-to-end latency were evaluated at a roadside unit receiver.

\hfill\break%
\noindent\textbf{Findings:} Dilithium-2 and SPHINCS+ exceeded the available transport-block capacity, whereas Falcon-512 remained physically feasible. Falcon-512 maintained mean latency near 52~ms and 95th-percentile latency within 97-98~ms, but met the 90\% packet-delivery threshold only at traffic level-of-service A, under line-of-sight propagation. ECDSA met the threshold through traffic level-of-service C. Neither algorithm met the threshold under non-line-of-sight propagation.

\hfill\break%
\noindent\textbf{Novelty:} The study provides a standards-grounded cross-layer evaluation that identifies both algorithm feasibility and traffic-dependent deployment boundaries for post-quantum signatures on C-V2X Mode 4 sidelink.

\hfill\break%
\noindent\textbf{Practical Applications:} The results show that spectrum efficiency, rather than cryptographic computation time, is the primary deployment constraint. They support standards development concerning payload structure, resource allocation, certificate transmission, and migration strategies for quantum-resistant vehicular communication.
\newpage
\section{Introduction}

The migration of vehicular security infrastructure toward PQC has transitioned from a long-term strategic consideration to a practical deployment priority with regulatory support. The U.S. National Security Agency's Commercial National Security Algorithm Suite 2.0 (CNSA 2.0) \cite{nsa_cnsa2} mandates the adoption of PQC algorithms in national security systems, with a full transition targeted for 2035. The timeline becomes especially critical in the context of Intelligent Transportation Systems (ITS), as vehicles deployed today are expected to operate for nearly 15 years, while roadside units (RSUs) may remain operational for up to 20 years \cite{bts_vehicle_age, fhwa_tms_lifecycle}. The expected lifespan of vehicles and RSUs deployed now overlaps with the projected timeframe in which a cryptographically relevant quantum computer (CRQC) may become operational \cite{nsa_cnsa2}. Beyond the immediate threat of signature forgery, a cryptographically relevant quantum computer (CRQC) would also introduce the threat of harvest-now-decrypt-later (HNDL) attacks \cite{mamun2026post}. In these attacks, adversaries record signed or encrypted transmissions today and decrypt or forge them once a CRQC becomes available. This makes timely migration necessary regardless of when such a quantum computer emerges. In vehicular communication, where transmitted messages include precise location, speed, and trajectory information, long-term data exposure poses significant privacy and safety concerns \cite{ieee_16092}.

Cellular Vehicle-to-Everything (C-V2X) technology, standardized by the 3rd Generation Partnership Project (3GPP) and deployed over the PC5 direct communication interface, constitutes the primary safety messaging platform for connected vehicles in the 5.9 GHz ITS band \cite{3gpp_36213, fcc2020cv2x}. V2X message security is governed by the IEEE 1609.2 standard \cite{ieee_16092}, which mandates ECDSA and a pseudonym certificate management framework executed through the Security Credential Management System (SCMS) \cite{brecht2018scms}. These ECDSA credentials are compact and well-suited to the sidelink channel's bandwidth constraints. However, the transition to PQC schemes produces authentication artifacts that are substantially larger. Whether NIST-standardized PQC digital signature algorithms (PQC-DSAs) remain compatible with existing protocol constraints without degrading safety-critical performance is a critical research question. The Society of Automotive Engineers (SAE) J3161 \cite{sae_j3161} deployment profile bounds the maximum payload transmissible in a single sidelink subframe through its transport block size (TBS) constraint. Because PQC signature and public key sizes far exceed those of ECDSA, some NIST-standardized schemes may be fundamentally incompatible with the current J3161 deployment profile, while others may impose measurable degradation on PDR and channel congestion. This study addresses these questions through theoretical TBS feasibility analysis and full-stack co-simulation of C-V2X Mode 4 sidelink across multiple traffic densities and channel propagation conditions. A transport block feasibility analysis eliminates Dilithium-2 and SPHINCS+ as physically incompatible with the SAE J3161 TBS ceiling under any permitted Modulation and Coding Scheme (MCS) configuration, leaving Falcon-512 as the only viable NIST PQC candidate for empirical evaluation.

IEEE 1609.2 requires every Basic Safety Message (BSM) to be signed individually. Because Mode 4 enables most distributed V2X safety communication without continuous cellular coverage, PQC overhead must first be evaluated on the Mode 4 sidelink before proposing revisions to existing standards. Therefore, this study evaluates PQC overhead on the BSM message type only. BSMs are broadcast at 10 Hz and constitute the highest-rate safety message defined in SAE J2735 \cite{sae_j2735}. They are also subject to the tightest J2945/1 latency and reliability requirements \cite{sae_j29451}. Other V2X message types, such as Signal Phase and Timing (SPaT), MAP, and Traveler Information Message (TIM), are RSU-originated, infrequent, and impose substantially lower aggregate sidelink loads. Evaluating PQC using BSMs therefore reflects the most demanding communication scenario, allowing the results to serve as a conservative upper bound for less demanding message types.

A simulation-based evaluation is necessary because no commercial on-board unit (OBU) or RSU platform currently supports PQC primitives in production firmware, so a fielded testbed cannot execute the algorithms under study. The evaluation also requires traffic density and channel propagation to be varied as controlled independent factors, which is not achievable in field experiments at densities of 50 to 100 veh/km. A standards-compliant co-simulation isolates the cryptographic algorithm as the only varying factor across scenarios. In a Mode 4 deployment, BSMs are broadcast on the PC5 sidelink and received by every neighboring vehicle and any RSU in range. The transmissions are not directed at a specific receiver, so vehicle-to-vehicle (V2V) and vehicle-to-infrastructure (V2I) receptions both occur on the same shared channel, and both contribute to aggregate sidelink contention. This study evaluates end-to-end latency and PDR at the RSU end. The RSU is a stationary observer with a known fixed position and a known profile to every vehicle, which provides a uniform measurement vantage point across all 24 scenarios and eliminates variations caused by changes in receiver geometry. 

Since prior research has examined PQC-V2X integration from hardware benchmarking, protocol adaptation, and implementation review perspectives, but no study has jointly characterized the sidelink communication impact of NIST-standardized PQC signature schemes across multiple traffic densities and propagation conditions within a standards-compliant C-V2X Mode 4 simulation, This study presents the first implementation and evaluation of NIST-standardized PQC-DSAs within a C-V2X PC5 Mode 4 sidelink simulation that maintains compliance with SAE J3161 deployment parameters and IEEE 1609.2 security standards, establishing both the feasible and infeasible operating regions for each algorithm. The following are the key contributions of this paper:

\begin{itemize}
    \item A feasibility assessment is conducted to determine which PQC algorithms can be accommodated within the current SAE J3161 sidelink deployment profile without modifications to the application layer, such as packet segmentation or hybrid cryptographic schemes, or to the radio resource configuration, such as MCS adaptation beyond the SAE J3161-specified range.
    
    \item This study characterizes the infeasibility of Dilithium-2 and SPHINCS+ under the existing SAE J3161 configuration. A cross-layer performance evaluation quantifies the impact of PQC signature and public key overhead on sidelink Key Performance Indicators (KPIs), including PDR and end-to-end latency, for both Falcon-512 and the ECDSA baseline under six traffic density levels spanning traffic level-of-service A through F.
 
    \item Under line-of-sight (LOS) propagation, the traffic density thresholds at which Falcon-512-induced overhead degrades sidelink performance below safety-critical requirements are identified, providing actionable deployment boundaries for standards development.
    
    \item Under non-line-of-sight (NLOS) propagation, the analysis shows that severe path loss renders both ECDSA and Falcon-512 unable to meet the 90\% PDR requirement at any communication distance, making propagation condition rather than algorithm choice the binding constraint.
\end{itemize}


\section{Related Work}
\label{sec:related}

C-V2X Mode 4 sidelink performance has been characterized through analytical modeling, simulation, and measurement. Gonzalez-Martin et al. \cite{gonzalez2019analytical} derived the first closed-form analytical models for Mode 4 PDR as a function of inter-vehicle distance, decomposing packet losses into half-duplex collisions, propagation failures, sensing-based scheduling collisions, and non-sensing collisions, with validation across vehicle densities. These models establish the baseline loss mechanisms that enlarged PQC payloads interact with. Bazzi et al. \cite{bazzi2017performance} compared IEEE 802.11p and Long Term Evolution (LTE)-V2V performance under highway scenarios and showed that C-V2X achieves superior range and reliability, providing a reference baseline for C-V2X sidelink behavior. Toghi et al. \cite{toghi2019congestion} analyzed Mode 4 under highly congested conditions and demonstrated the role of CBR-based congestion control in maintaining channel stability at high vehicle densities. Collectively, these works characterize baseline C-V2X PC5 Mode 4 behavior under compact classical cryptographic payloads but do not account for the impact of large PQC authentication artifacts on sidelink resource allocation or communication KPIs.

The IEEE 1609.2 standard defines V2X security services, including ECDSA-based digital signatures and a pseudonym certificate management framework. Brecht et al. \cite{brecht2018scms} detailed the SCMS architecture for V2X certificate management. The architecture was designed around the compact ECDSA P-256 credentials, leaving open the question of how SCMS scales when larger PQC credentials replace this baseline. Qu et al.~\cite{qu2015security} reviewed security and privacy challenges in Vehicular Ad Hoc Networks (VANETs) and identified authentication efficiency, signature verification delay, and computational overhead as major challenges for secure vehicular communications. Lu et al.~\cite{lu2018survey_v2x_security} reviewed security and privacy frameworks for 5G V2X and emphasized that security and authentication mechanisms must satisfy the stringent latency and reliability requirements of safety-critical vehicular services. While these works address V2X security architecture comprehensively, they do not evaluate the radio resource impact of PQC alternatives on the C-V2X sidelink communication.

Hardware benchmarking studies have confirmed PQC computational feasibility on automotive hardware platforms, but leave the communication-layer impact unaddressed. Wang and St\"{o}ttinger \cite{wang2020automotive} demonstrated Field-Programmable Gate Array (FPGA)-based hardware accelerators for post-quantum automotive Hardware Security Modules (HSMs), establishing the viability of PQC integration in automotive platforms. Howe and Westerbaan \cite{howe2023benchmarking} benchmarked CRYSTALS-Dilithium and Falcon on the ARM Cortex-M7 and showed that Falcon achieves 6 to 8 times faster signing through native 64-bit Floating-Point Unit (FPU) acceleration of its floating-point-intensive operations. Sinell et al.~\cite{sinell2025evaluation} evaluated Falcon on the Infineon AURIX TC399 automotive multicore processor and confirmed that verification latency meets real-time V2X requirements, although key generation exhibited high latency variance that challenges predictable real-time operation. Mishra~\cite{sae2026pqc_twowheeler} evaluated NIST PQC algorithms on ARM Cortex-M microcontrollers representative of two-wheeler Electronic Control Units (ECUs), benchmarking signing and verification latency as well as memory footprint, and demonstrated the feasibility of deploying lattice-based PQC schemes on constrained embedded platforms. None of these studies models how the enlarged authentication payloads produced by PQC schemes propagate through C-V2X Mode 4 sidelink scheduling, congestion control, and physical-layer resource allocation. At the protocol and system level, Fernandez-Carames and Fraga-Lamas \cite{fernandez2020towards} reviewed PQC approaches for blockchain-based security systems, identifying lattice-based schemes, particularly Falcon, as promising candidates due to their compact signature sizes relative to other PQC families. Barreto et al. \cite{barreto2018qscms} extended the SCMS certificate management architecture to lattice-based primitives, demonstrating that a PQC-compatible certificate infrastructure is architecturally feasible. Lonc et al. \cite{lonc2023feasibility} benchmarked NIST PQC candidates against European Telecommunications Standards Institute Cooperative ITS (ETSI C-ITS) bandwidth and latency budgets and identified viable replacements for Elliptic Curve Cryptography (ECC), but without integrating the selected algorithms into a C-V2X Mode 4 environment. Bindel et al. \cite{bindel2017transitioning} established security proofs for classical-PQC composite signatures, providing a theoretical basis for hybrid vehicular certificate designs, but without addressing the bandwidth and timing constraints of broadcast vehicular safety messaging.

The most directly relevant works examine how large PQC payloads interact with V2X protocol constraints. Kim and Seo \cite{kim2023splitmethod} proposed splitting PQC signatures across consecutive BSM transmissions to maintain IEEE 1609.2 compliance without modifying the transport block structure. Twardokus et al. \cite{twardokus2024hand} introduced a certificate scheduling technique that exploits redundancy in V2V certificate transmissions to fit NIST PQC schemes within the Dedicated Short Range Communications (DSRC) spectrum budget, validated using a software-defined radio (SDR) testbed, but the approach targets DSRC and does not address C-V2X Mode 4's autonomous resource reselection mechanism or the TBS constraints imposed by SAE J3161. Chen et al. \cite{chen2024hybrid} validated a hybrid Falcon-ECC pseudonym certificate scheme through OBU field tests in a live C-ITS deployment. Twardokus and Rahbari \cite{twardokus2025chasm} introduced CHASM, a cross-layer framework combining certificate fragmentation with adaptive modulation to reduce per-packet sidelink overhead for PQC certificates in C-V2X. While CHASM demonstrates that cross-layer design can partially offset PQC size penalties for a single algorithm, it does not assess schemes that exceed the sidelink TBS ceiling entirely, does not evaluate performance across multiple traffic density levels, and does not model the interaction between larger PQC payloads and Mode 4's autonomous resource allocation.

\subsection{Research Gap}

Across all four research directions, no prior work jointly characterizes how the NIST-standardized PQC signature algorithms perform across multiple traffic density levels and both LOS and NLOS propagation conditions within a single, standards-compliant C-V2X Mode 4 simulation spanning the full protocol stack from application-layer payload generation through Medium Access Control (MAC)-layer resource scheduling to physical-layer transmission. Prior studies have not determined when increasing traffic density degrades sidelink performance due to PQC overhead, how propagation conditions affect PQC-loaded channels, or which NIST-standardized schemes remain feasible under SAE J3161 constraints. This study addresses these gaps.

\section{Preliminaries}
\label{sec:background}

\subsection{C-V2X Mode 4 and SAE J3161}
\label{sec:cv2x}

C-V2X is a 3GPP-standardized radio access technology that enables direct V2V and V2I communication over the PC5 sidelink interface without requiring cellular network infrastructure \cite{3gpp_36213}. The PC5 interface operates in the 5.9 GHz ITS band and supports two transmission modes: Mode 3, in which an infrastructure called evolved Node B (eNB) schedules sidelink resources centrally; and Mode 4, in which vehicles and RSUs autonomously select resources using a sensing-based Semi-Persistent Scheduling (SPS) mechanism \cite{hajisami2022tutorial}. Mode 4 is the deployment-relevant mode for safety-critical V2X for two reasons. First, it operates without cellular infrastructure, which makes it the only viable option in coverage-edge environments such as rural roads, tunnels, and unserved intersections where eNB scheduling is unavailable. Second, the SAE J3161 deployment profile is defined exclusively against Mode 4 operation \cite{hajisami2022tutorial, sae_j3161}. This study evaluates PQC on the Mode 4 sidelink, where the impact of larger authentication payloads is most consequential for real-world deployment. In Mode 4, each vehicle monitors the sidelink spectrum over a 1,000 ms sensing window and identifies candidate single-subframe resources within a selection window. Resources whose measured Reference Signal Received Power (RSRP) exceeds a threshold are excluded from the candidate set. The vehicle then randomly selects a resource from the remaining candidates and reserves it for a configurable Resource Reservation Interval (RRI). A reselection counter determines how many consecutive transmissions use the reserved resource before triggering a new selection. At each reselection event, the vehicle retains the current resource with a probability $p_\text{keep}$ (typically 0.0, 0.2, 0.4, 0.6, or 0.8) or selects a new resource from the updated candidate set. Sidelink payloads are carried on the Physical Sidelink Shared Channel (PSSCH), while the associated Sidelink Control Information (SCI) is transmitted on the Physical Sidelink Control Channel (PSCCH). The SCI carries scheduling metadata---including the MCS index, subchannel allocation, and resource reservation offset---that receivers need to locate and decode the accompanying PSSCH transport block. Adjacent PSCCH-PSSCH mapping places control and data subchannels in contiguous resource blocks within the same subframe. The MCS applied to the PSSCH determines the TBS and, consequently, the maximum payload that can fit within a given subchannel allocation \cite{3gpp_36321, hajisami2022tutorial, gonzalez2019analytical}.

SAE J3161 \cite{sae_j3161} specifies the deployment profile for C-V2X V2V communication in the 5.9 GHz band. It mandates a 20 MHz channel bandwidth corresponding to 100 resource blocks (RBs) per subframe, a subchannel size of 10 RBs yielding 10 subchannels per subframe, and an RRI of 100 ms aligned with the BSM generation rate defined in SAE J2945/1. The standard further defines congestion control through Channel Busy Ratio (CBR) monitoring and restricts the allowed MCS indices and maximum subchannel allocations as a function of measured CBR. At the most permissive CBR threshold, MCS 11 with a maximum allocation of 10 subchannels yields a TBS ceiling of 2,481 bytes. This bound directly constrains the maximum payload that the sidelink can carry and is the key feasibility limit for PQC adoption.

\subsection{V2X Security and Post-Quantum Cryptography}
\label{sec:pqc}

V2X message authentication is governed by IEEE 1609.2~\cite{ieee_16092}, which defines security services for Wireless Access in Vehicular Environments (WAVE) application and management messages. The standard specifies a Secure Data Service that wraps each transmitted message in a Services Protocol Data Unit (SPDU). A signed SPDU provides four security properties: authenticity, which assures that the sender is who they claim to be; authorization, which assures that the sender possesses the permissions required for the message type; integrity, which ensures that any modification after signing can be detected; and non-repudiation of origin. The signer's authorization to transmit a specific message type is established through application permissions encoded in the certificate using Provider Service Identifiers (PSIDs), which may be further scoped to specific geographic regions and validity periods. Each SPDU carries a \texttt{SignerIdentifier} field containing either the signer's full pseudonym certificate or a compact 8-byte \texttt{HashedId8} digest. An SPDU carrying the full certificate is referred as the certificate-bearing SPDU, and the corresponding transmission configuration is referred as a "certificate mode" in this study. An SPDU carrying only the digest is referred as a digest-bearing SPDU, and its configuration is referred as "digest mode." Full certificates include the verification public key, application permissions, a geographic validity region, a validity period, and the certificate authority's signature on the certificate. To protect vehicle operator privacy, IEEE 1609.2 supports pseudonym certificate rotation. Each vehicle holds a pool of concurrently valid pseudonym certificates and periodically changes the active certificate to prevent an outside observer from linking successive transmissions to a specific vehicle. Revoked certificates are distributed through Certificate Revocation Lists (CRLs) using linkage-based identifiers, which enable efficient revocation without revealing certificate associations before revocation occurs. The standard mandates ECDSA with the NIST P-256 curve as the digital signature algorithm. An ECDSA P-256 signature is 64 bytes, and the corresponding public key is 65 bytes, both of which fit comfortably within a sidelink transport block.

The anticipated CRQC could run Shor's algorithm~\cite{shor1994algorithms} to solve the elliptic curve discrete logarithm problem efficiently, breaking ECDSA. NIST responded with a multi-year standardization effort~\cite{nist_pqc_2024} and published three PQC digital signature standards in August 2024, each built on computational problems believed to be intractable for both classical and quantum adversaries. NIST evaluates the strength of PQC algorithms using five security levels. Level 1 is the minimum, targeting security equivalent to a brute-force key search on Advanced Encryption Standard (AES)-128. Level 2 is stronger, targeting security equivalent to finding a collision in Secure Hash Algorithm (SHA)-256. Module-Lattice-Based Digital Signature Algorithm (ML-DSA), standardized as Federal Information Processing Standard (FIPS) 204~\cite{fips204} and derived from CRYSTALS-Dilithium, is a lattice-based scheme whose security rests on the hardness of the Module Learning With Errors (MLWE) problem. ML-DSA-44, also known as Dilithium-2, meets NIST security Level 2 and produces 2,420-byte signatures with a 1,312-byte public key. The signing process is deterministic, which simplifies integration, but the public key and signature sizes are substantially larger than ECDSA. The Stateless Hash-Based Digital Signature Algorithm (SLH-DHA) is derived from SPHINCS+ and standardized as FIPS 205~\cite{fips205}, relies solely on the security of cryptographic hash functions, and introduces no new algebraic hardness assumptions. At NIST security level 1, the SPHINCS+-SHA2-128s parameter set produces 7,856-byte signatures with a 32-byte public key. Its signatures are the largest among the three standardized schemes, making it the most challenging candidate for bandwidth-constrained deployment. Falcon~\cite{fouque2017falcon} is a lattice-based scheme constructed over N-th degree Truncated Polynomial Ring Units (NTRU) lattices that uses Fast Fourier Sampling for efficient signing. NIST has selected Falcon for standardization as FIPS 206, with finalization pending as of this writing~\cite{nist_pqc_2024}. At NIST security level 1, the Falcon-512 parameter set produces a 666-byte signature with a 897-byte public key. This compact profile makes Falcon-512 the most bandwidth-efficient among the three NIST-selected PQC signature schemes, though even its signature is approximately ten times larger than the ECDSA counterpart~\cite{lonc2023feasibility}. Table~\ref{tab:crypto_sizes} summarizes the signature and public key sizes for all NIST-standardized PQC digital signature algorithms along with the ECDSA baseline. The disparity between ECDSA and the PQC schemes establishes the central research and deployment challenge addressed in this paper: each additional byte in a secured BSM payload must be transmitted through the shared sidelink channel, where the TBS is bounded by the SAE J3161 deployment profile.

\begin{table}[!htbp]
      \caption{Cryptographic Algorithm Parameter Sizes}
      \label{tab:crypto_sizes}
      \centering
      \small
      \resizebox{\textwidth}{!}{%
      \begin{tabular}{lccc}
      \toprule
      \textbf{Algorithm} & \textbf{Signature (Bytes)} & \textbf{Public Key (Bytes)} & \textbf{NIST Security Level} \\
      \midrule
      ECDSA P-256              & 64    & 65    & Classical \\
      Falcon-512       & 666   & 897   & 1         \\
      ML-DSA-44 (Dilithium-2)  & 2,420 & 1,312 & 2         \\
      SLH-DSA-SHA2-128s (SPHINCS+-SHA2-128s)       & 7,856 & 32    & 1         \\
      \bottomrule
      \end{tabular}%
      }
  \end{table}

\section{Method}
\label{sec:method}

\begin{figure}[t]
\centering
\resizebox{0.94\textwidth}{!}{
\begin{tikzpicture}[
  font=\footnotesize,
  node distance=8mm and 16mm,
  basebox/.style={
    draw=gray!30,
    thick,
    rounded corners=3mm,
    align=center,
    inner sep=6pt,
    blur shadow={shadow blur steps=5, shadow xshift=1pt, shadow yshift=-1pt}
  },
  circlebox/.style={basebox, fill=setupblue, draw=darkblue, text width=2.4cm, minimum height=1.6cm},
  corebox/.style={basebox, fill=setupblue, draw=darkblue, text width=4.2cm},
  elimbox/.style={basebox, fill=excludered!30, draw=excludered, text width=3.8cm},
  headerorange/.style={basebox, fill=orange!80!black, draw=orange!80!black, text width=6.2cm, font=\footnotesize\bfseries, text=white},
  childorange/.style={basebox, fill=processorange, draw=orange!80!black, text width=2.7cm},
  headergreen/.style={basebox, fill=green!45!black, draw=green!45!black, text width=5.6cm, font=\footnotesize\bfseries, text=white},
  childgreen/.style={basebox, fill=resultgreen, draw=green!55!black, text width=4.3cm},
  arrow/.style={-{Latex[length=3mm, width=2mm]}, thick, darkblue!80},
  dashedarrow/.style={arrow, dashed, gray}
]
 
\node[circlebox] (sae) {\textbf{\textcolor{darkblue}{SAE J3161}}\\Requirements};
\node[circlebox, below=10mm of sae] (ieee) {\textbf{\textcolor{darkblue}{IEEE 1609.2}}\\Requirements};
 
\node[corebox] (core) at ($(sae)!0.5!(ieee) + (5.0cm,0)$) {
    \textbf{\textcolor{darkblue}{Implementation Feasibility Analysis}} of NIST-standardized PQC-DSAs in C-V2X PC5 Mode 4
};
 
\node[elimbox, below=12mm of core] (elim) {
    \textbf{\textcolor{excludered}{Dilithium-2 and SPHINCS+}} eliminated for simulation
};
 
\node[headerorange, right=18mm of core] (setuphead) {Simulation Setup};
\node[childorange, below=5mm of setuphead, xshift=-1.72cm] (lib) {PQC Library Integration OpenCV2X};
\node[childorange, right=3mm of lib] (traffic) {Traffic Scenario Generation in SUMO for 6 traffic levels-of-service};
 
\node[headergreen, right=18mm of setuphead] (loghead) {Simulation Log Post Processing};
\node[childgreen, below=5mm of loghead] (pdr) {PDR Calculation};
\node[childgreen, below=2mm of pdr] (latency) {End-to-End Latency Calculation};
 
\begin{pgfonlayer}{background}
    \node[draw=orange!50, fill=processorange!30, rounded corners, fit=(setuphead)(lib)(traffic), inner sep=4mm] (setupgroup) {};
    \node[draw=green!50!black, fill=resultgreen!30, rounded corners, fit=(loghead)(pdr)(latency), inner sep=4mm] (loggroup) {};
\end{pgfonlayer}
 
\draw[arrow] (sae.east) -- (core.north west);
\draw[arrow] (ieee.east) -- (core.south west);
\draw[arrow] (core.south) -- (elim.north);
\draw[arrow] (core.east) -- (setuphead.west);
\draw[arrow] (setuphead.east) -- (loghead.west);
 
\end{tikzpicture}
}
\caption{Overview of the evaluation method: SPDU sizes computed from NIST PQC algorithm parameters and IEEE 1609.2 SPDU format. The SAE J3161 channel profile determines each algorithm's feasibility. Feasible PQC candidates and baseline ECDSA are then evaluated across 24 co-simulation scenarios spanning six traffic density levels and two propagation conditions.}
\label{fig_pqc_methodology}
\end{figure}
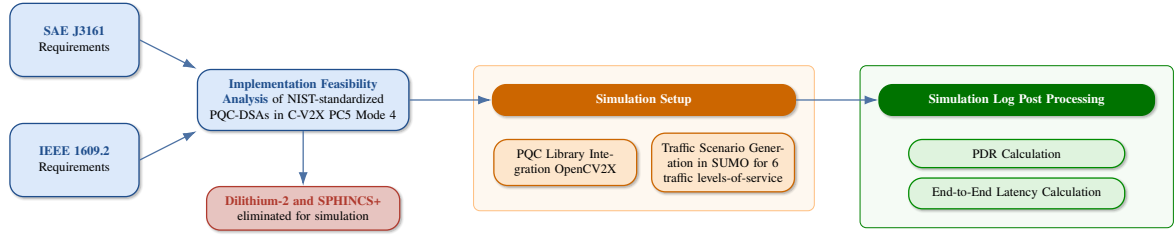

This section describes the method used to evaluate PQC authentication on C-V2X Mode 4 sidelink. The evaluation proceeds in two stages. First, a transport block feasibility analysis determines which NIST-standardized PQC algorithms are physically compatible under the constraint of the SAE J3161 MCS value. This analysis drives algorithm selection: Dilithium-2 and SPHINCS+ are eliminated because their SPDU sizes exceed the maximum transport block capacity under any permitted MCS and subchannel configuration. Falcon-512 is the only NIST PQC scheme that fits within the J3161 profile and is therefore the sole PQC candidate carried forward to empirical evaluation. Second, Falcon-512 and ECDSA P-256 are compared under identical C-V2X Mode 4 conditions across six traffic density levels and two channel configurations, yielding 24 total simulation scenarios. In each scenario, vehicular mobility, the channel model, and radio resource configuration are held constant, so any observed performance difference can be attributed directly to the change in cryptographic algorithm and payload size. The following subsections describe the SAE J3161 deployment profile, the sidelink resource allocation feasibility analysis, the IEEE 1609.2 security profile, the co-simulation platform, the traffic scenario configuration, the channel propagation models, and the KPI definitions used for evaluation. Figure~\ref{fig_pqc_methodology} summarizes the overall evaluation method.

\subsection{SAE J3161 Deployment Profile}
\label{sec:j3161_profile}

The radio parameters in this study follow the SAE J3161 \cite{sae_j3161} deployment profile for LTE-V2X PC5 sidelink communication. The standard specifies a 20 MHz channel bandwidth centered at 5915 MHz, providing 100 resource blocks per 1 ms subframe. Subchannels are formed with a fixed size of 10 RBs, yielding 10 subchannels per subframe. Adjacent PSCCH-PSSCH mapping places the control and data channels in contiguous resource blocks within the same subframe. SAE J3161 defines two PSSCH transmission parameter sets, selected by a speed threshold of 120 km/h. Since all vehicles in this study operate at a free-flow speed of 72.4 km/h, the low-speed parameter set applies. Under this set, the valid MCS indices are 5, 6, 7, and 11; indices 8, 9, and 10 are explicitly excluded by the standard. MCS 5, 6, and 7 apply QPSK modulation with increasing code rates, while MCS 11 applies 16-QAM. The minimum subchannel allocation is 2, and the maximum is 10. For SPS transmission at low speed, J3161 Table 16 maps packet size to (MCS, RB) pairs; at MCS 11 with the full 10-subchannel allocation, the maximum transport block that can be carried without segmentation or dropping is 2,481 bytes. CBR-based congestion control is defined over three zones: CBR $<$ 0.30, $0.30 \leq$ CBR $<$ 0.65, and CBR $\geq$ 0.65. Within each zone, a channel occupancy ratio (CR) limit is applied per ProSe per-packet priority (PPPP). For BSMs transmitted as essential V2V traffic (PPPP 5), the limits are 8\%, 3\%, and 1.5\%. Vehicles transmit at 23 dBm, and the resource keep probability is set to 0.8, as specified in the standard. SAE J3161 Section 7.3.3 requires SPS+One-Shot transmission, in which a periodic one-shot packet is interleaved with the SPS flow to detect and break repetitive scheduling collisions caused by half-duplex operation. The OpenCV2X simulator does not implement this mechanism. Hybrid Automatic Repeat Request (HARQ) retransmissions are permitted by the standard but are disabled in this simulation. Disabling HARQ isolates first-transmission resource allocation as the sole independent variable, ensuring that observed performance differences reflect only packet size and not retransmission scheduling.

\subsection{IEEE 1609.2 Security Profile}
\label{sec:packet_structure}

Each transmitted packet follows the IEEE~1609.2 SPDU structure~\cite{ieee_16092}. The SPDU encapsulates a J2735 BSM Part I (BSMcoreData), which contains vehicle position, speed, heading, and other kinematic data totaling 39~bytes per the J2735 standard~\cite{sae_j2735}. The IEEE 1609.2 \texttt{Ieee1609Dot2Data} wrapper adds a protocol version, content type, hash algorithm identifier, PSID, generation time, generation location, and encoding overhead, contributing approximately 28 bytes of fixed overhead. The \texttt{SignerIdentifier} field follows the pseudonym certificate inclusion policy specified in SAE J2945/1 \cite{sae_j29451}:  the vehicle transmits in certificate mode for every fifth BSM, embedding the full explicit certificate, while the remaining four transmissions use digest mode, carrying only the 8-byte \texttt{HashedId8} digest. When the full certificate is included, it comprises a fixed overhead of approximately 105 bytes, including version, type, issuer identifier, application permissions, validity period, and verification key indicator, in addition to the raw public key of the signing algorithm. The total SPDU size for each algorithm when transmitting with a full certificate is:
\begin{equation}
    S_\text{SPDU} = S_\text{header} + S_\text{BSM} + S_\text{sig} + S_\text{cert}
    \label{eq:spdu_size}
\end{equation}
where $S_\text{header} = 28$ bytes is the IEEE 1609.2 header overhead, $S_\text{BSM} = 39$ bytes is the BSM payload, $S_\text{sig}$ is the digital signature size, and $S_\text{cert}$ is the full certificate size. If $S_\text{pk}$ is the raw public key size, then $S_\text{cert} = 105 + S_\text{pk}$. Table~\ref{tab:spdu_sizes} presents the resulting SPDU sizes for all four algorithms evaluated in this study.

\begin{table}[!htbp]
    \caption{SPDU Sizes with Full Certificate}
    \label{tab:spdu_sizes}
    \centering
    \small
    \begin{tabular}{lccc}
    \toprule
    \textbf{Algorithm} &
    \textbf{Sig (Bytes)} &
    \textbf{Cert (Bytes)} &
    \textbf{Total SPDU (Bytes)} \\
    \midrule
    ECDSA P-256         & 64    & 170   & 301   \\
    Falcon-512          & 666   & 1,002 & 1,735 \\
    Dilithium-2         & 2,420 & 1,417 & 3,904 \\
    SPHINCS+-SHA2-128s  & 7,856 & 137   & 8,060 \\
    \bottomrule
    \end{tabular}
\end{table}

As established earlier, the Mode 4 SPS grant is allocated at resource selection time for the certificate-bearing SPDU, making it the binding feasibility constraint regardless of how many transmissions within the reservation period actually carry the full certificate. All five transmissions within a reservation period occupy the same number of subchannels regardless of whether a full certificate is included in any individual transmission. The certificate-bearing SPDU size is the determining constraint for sidelink feasibility analysis.

\subsection{Sidelink Resource Allocation Feasibility}
\label{sec:resource_feasibility}

The C-V2X sidelink TBS determines the maximum payload that can be transmitted in a single subframe. The TBS is a function of the number of allocated physical resource blocks (PRBs) and the MCS index, as specified in 3GPP TS 36.213 Table 7.1.7.2.1-1 \cite{3gpp_36213}. For adjacent PSCCH-PSSCH resource mapping with a subchannel size of $N_\text{RB}^\text{sc}$ RBs and $N_\text{sc}$ allocated subchannels, the number of PSSCH PRBs available for data is:
\begin{equation}
    N_\text{PRB}^\text{PSSCH} = N_\text{sc} \cdot N_\text{RB}^\text{sc} - 2
    \label{eq:nprb}
\end{equation}
where 2 PRBs are reserved for the PSCCH carrying the SCI. The MCS index determines the transport block size index $I_\text{TBS}$ and the modulation order $Q_m$ through the MCS-to-TBS mapping table \cite{3gpp_36213}. The TBS in bits is then obtained by looking up $I_\text{TBS}$ and $N_\text{PRB}^\text{PSSCH}$ in the TBS table.

Under the SAE J3161 deployment profile, the MCS range is restricted to indices 5,6,7, and 11, and the maximum subchannel allocation is 10 subchannels with 10 RBs each, yielding $N_\text{PRB}^\text{PSSCH} = 98$. At MCS 11 ($I_\text{TBS} = 10$, 16QAM), the maximum TBS with 10 subchannels is 19,848 bits (2,481 bytes). Table \ref{tab:tbs_feasibility} presents the subchannel requirements for each algorithm at representative MCS levels. The analysis reveals several critical feasibility constraints. Dilithium-2 SPDUs with full certificates (3,904 bytes) exceed the maximum TBS at any MCS within the J3161-permitted range, even when all 10 subchannels are allocated. Moreover, Dilithium-2's digest-bearing SPDU of 2,495 bytes still exceeds the maximum TBS of 2,481 bytes, an infeasibility that persists regardless of the fact that the full certificate is transmitted once in every 5 BSM. SPHINCS+-SHA2-128s, with a minimum signature size of 7,856 bytes, is even further from feasibility. Accommodating these algorithms without certificate segmentation would require protocol modifications beyond the current J3161 profile, which is out of scope of this study.

\begin{table*}[!t]
    \caption{Minimum Subchannels Required per Algorithm and MCS}
    \label{tab:tbs_feasibility}
    \centering
    \small
    \begin{tabular}{llcccc}
        \toprule
        \textbf{Algorithm} &
        \textbf{Mode} &
        \textbf{SPDU Size} &
        \multicolumn{3}{c}{\textbf{Subchannels Required}} \\
        \cmidrule(lr){4-6}
        & & \textbf{(B)} &
        \textbf{MCS 5} &
        \textbf{MCS 7} &
        \textbf{MCS 11} \\
        \midrule
        ECDSA       & Cert   & 301     & 3   & 3   & 2   \\
        ECDSA       & Digest & 139     & 2   & 2   & 1   \\
        Falcon-512  & Cert   & 1{,}739 & N/A & N/A & 7   \\
        Falcon-512  & Digest & 741     & 7   & 5   & 4   \\
        Dilithium-2 & Cert   & 3{,}904 & N/A & N/A & N/A \\
        Dilithium-2 & Digest & 2{,}495 & N/A & N/A & N/A \\
        \bottomrule
    \end{tabular}
\end{table*}

Falcon-512 occupies a narrow but viable operating region. Certificate-bearing SPDUs (1,735 bytes) require MCS 11 with 7 subchannels, consuming 70\% of the available sidelink resources for a single transmission. Digest-bearing SPDUs of 741 bytes are more resource-efficient, requiring only 4 subchannels at MCS 11. This infeasibility extends to all higher-security parameter sets of the standardized schemes. Falcon-1024 with the NIST security level 5 produces a 1,280-byte signature with a 1,793-byte public key, yielding a certificate-bearing SPDU of approximately 3,245 bytes, which is well above the TBS ceiling. ML-DSA-65 and ML-DSA-87, the NIST Level 3 and Level 5 variants of ML-DSA, produce signatures of 3,309 and 4,627 bytes, respectively, and are infeasible by even larger margins. All higher-security parameter sets of SPHINCS+ similarly produce larger signatures than the Level 1 variant already eliminated above. Falcon-512 at NIST Level 1 is therefore the only parameter set, across all security levels of all three standardized schemes, that fits within the SAE J3161 profile. This feasibility analysis motivates the focus of the evaluation on Falcon-512 as the primary PQC candidate alongside the ECDSA baseline. 

An important consequence of Mode 4 SPS scheduling is how the full certificate transmission policy interacts with MAC layer grant allocation. Under the SAE J2945/1, four out of every five BSMs carry a compact 8-byte HashedId8 digest rather than the full certificate. When the first BSM carrying the full certificate is generated, the initial SPS grant allocates 7 subchannels. When this grant expires and a new resource is selected via SPS reselection, the probability that the triggering packet is a digest is 80\%, resulting in a 4-subchannel grant in most cases. Once a 4-subchannel grant is established, certificate-bearing BSMs that arrive on it do not trigger a grant regeneration. Instead, the RLC Unacknowledged Mode (UM) layer fragments the oversized packet: it sends what fits within the current period's transport block and defers the remainder to the next transmission period. Because RLC UM provides no retransmission, both fragments must be received intact for the BSM to be reassembled at the receiver; losing either fragment permanently discards the entire BSM. The system therefore settles into a steady state of 4-subchannel grants, which is reflected in the mean number of subchannels used being 4 across all Falcon simulation scenarios. This behavior has two important implications. First, for PQC feasibility on Mode 4 sidelink, steady-state channel allocation is governed by the digest SPDU size: Falcon's 4-subchannel grant occupies 40\% of the sidelink resource pool per transmission. However, the certificate transmission itself triggers RLC fragmentation and elevates loss probability, introducing a reliability penalty specific to PQC certificate exchanges that is absent in ECDSA, whose certificates fit within a single transport block. Second, this further reinforces the infeasibility of Dilithium-2: even the digest-bearing SPDU, at 2,495 bytes, exceeds the maximum TBS of 2,481 bytes, an infeasibility that persists regardless of the certificate refresh interval.

\subsection{Co-Simulation Platform}
\label{sec:sim_framework}

The co-simulation framework integrates an LTE PC5 Mode 4 network simulator with a microscopic traffic simulator to capture the interaction between vehicular mobility and sidelink radio resource allocation. The PC5 Mode 4 network simulation uses OpenCV2X \cite{opencv2x}, an open-source extension of SimuLTE \cite{virdis2014simulte} built on the Objective Modular Network Testbed in C++ (OMNeT++) discrete-event simulation platform \cite{varga2019omnetpp}. OpenCV2X models the full LTE protocol stack, including the physical layer, MAC-layer Mode 4 SPS resource allocation, Radio Link Control (RLC), and Packet Data Convergence Protocol (PDCP) layers. OpenCV2X is coupled with Vehicles in Network Simulation (Veins) \cite{sommer2019veins} for vehicular networking abstractions and with the Simulation of Urban Mobility (SUMO) \cite{lopez2018sumo} traffic simulator through the Traffic Control Interface (TraCI), enabling co-simulation of vehicular communication and realistic microscopic mobility. PQC algorithms are implemented using the liboqs library \cite{stebila2020liboqs}, which provides reference implementations of NIST-standardized PQC algorithms. ECDSA P-256 is implemented using OpenSSL \cite{openssl}. Each vehicle and RSU generates a fresh key pair at initialization and performs real-time signing and verification of every BSM, capturing actual cryptographic overhead in the simulation timing. In a Mode 4 deployment, BSMs are broadcast on the PC5 sidelink and received by every peer vehicle and any RSU in range. This study evaluates PDR and end-to-end latency at the RSU, which serves as a stationary observer with a known fixed position and a known propagation profile to every vehicle, providing a uniform measurement vantage point across all 24 scenarios. The V2V receptions that occur in parallel are not logged for analysis, but the corresponding transmissions are fully simulated and contribute to the aggregate channel load. Table~\ref{tab:sim_params} summarizes the complete set of simulation parameters, where radio parameters are configured according to SAE J3161.

\begin{table}[!htbp]
    \caption{Simulation Parameters}
    \label{tab:sim_params}
    \centering
    \small
    \begin{tabular}{lc}
    \toprule
    \textbf{Parameter} & \textbf{Value} \\
    \midrule
    Carrier frequency                          & 5915 MHz \\
    Channel bandwidth                          & 20 MHz \\
    Number of resource blocks                  & 100 per subframe \\
    Subchannel size                            & 10 RBs \\
    Number of subchannels                      & 10 \\
    MCS indices (low speed, per J3161)         & 5, 6, 7, 11 \\
    Transmit power                             & 23 dBm \\
    Resource reservation interval (RRI)       & 100 ms \\
    Resource keep probability ($p_\text{keep}$) & 0.8 \\
    BSM generation rate                        & 10 Hz (100 ms) \\
    Simulation duration                        & 300 s \\
    Congestion control                         & CBR-based with CR limiting \\
    HARQ retransmissions                       & disabled \\
    \bottomrule
    \end{tabular}
\end{table}

\subsection{Traffic Scenario}
\label{sec:traffic_scenario}

The traffic scenario models a four-leg signalized urban intersection with two lanes per approach, generated using SUMO. The four-leg signalized intersection is selected as the evaluation geometry for several reasons. First, it is the most common intersection configuration in the U.S. urban road network \cite{fhwa_sig_guide}, representing the standard deployment environment for C-V2X safety applications \cite{sae_j3161}. Second, a four-leg intersection produces 32 potential vehicle-vehicle conflict points, of which 16 are crossing conflicts where vehicle paths intersect at near-right angles \cite{fhwa_sig_guide}. These crossing conflicts are the primary target of V2X safety applications, such as intersection collision avoidance, making the four-leg geometry the most safety-critical intersection type for evaluating authentication reliability. Third, vehicles arriving from all four approaches converge within mutual communication range simultaneously, producing the highest concentration of concurrent sidelink transmitters and thereby the most demanding test of Mode 4 resource allocation under PQC-induced payload overhead. Additionally, vehicles on perpendicular approaches experience NLOS propagation due to corner building obstruction, enabling evaluation under both LOS and NLOS channels within a single scenario.

\begin{figure}[t]
    \centering
    \includegraphics[width=\textwidth]{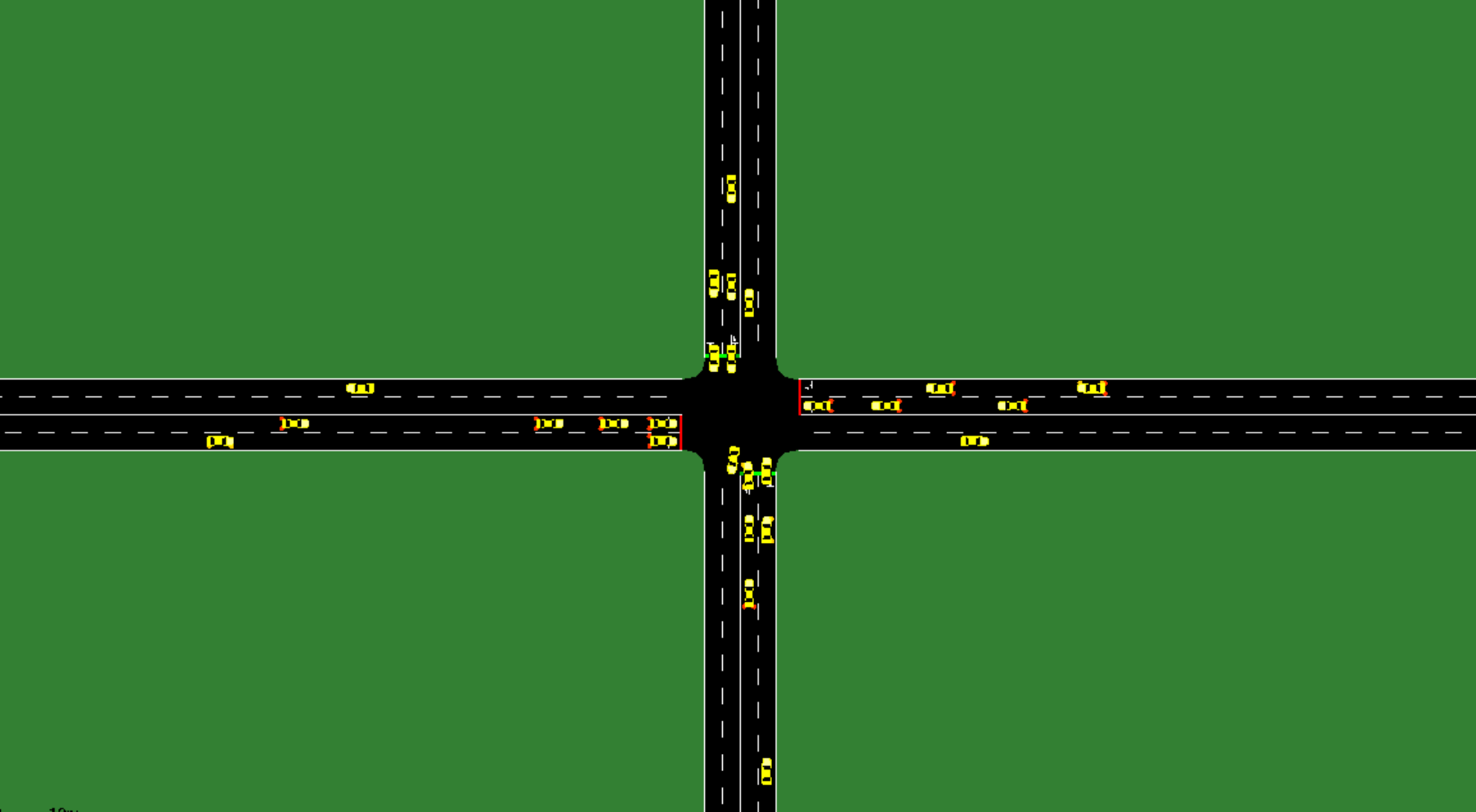}
    \caption{SUMO-generated four-leg signalized intersection with two lanes per approach and an 80/10/10 through-left-right movement split.}

    \label{fig:sumo}
\end{figure}

Figure~\ref{fig:sumo} shows the simulated intersection. Vehicles follow the Krauss car-following model in SUMO with an acceleration of 2.6 m/s\textsuperscript{2}, a deceleration of 4.5 m/s\textsuperscript{2}, a vehicle length of 5 m, and a free-flow speed of $v_f = 20.12$ m/s (72.4 km/h). Traffic is distributed with an 80/10/10 split among through, left-turn, and right-turn movements on each approach, reflecting typical urban intersection movement proportions. Six traffic density levels, level-of-service A through F, following the Highway Capacity Manual (HCM) arterial segment density classification \cite{HCM2016}, are evaluated to characterize PQC sidelink performance across the full range of operating conditions from free-flow to severely oversaturated traffic. The fundamental relationship between traffic flow rate $q$ (veh/h), density $k$ (veh/km), and space mean speed $v$ (km/h) is given by:
\begin{equation}
    q = k \cdot v
    \label{eq:flow_density}
\end{equation}

\begin{table}[!htbp]
    \caption{Traffic Density Scenarios}
    \label{tab:traffic_scenarios}
    \centering
    \small
    \begin{tabular}{lcc}
        \toprule
        \textbf{Traffic Level of Service} &
        \textbf{Density (veh/km)} &
        \textbf{Flow per Approach (veh/h)} \\
        \midrule
        A & 10  & 724   \\
        B & 20  & 1{,}448 \\
        C & 30  & 2{,}172 \\
        D & 40  & 2{,}896 \\
        E & 50  & 3{,}620 \\
        F & 100 & 7{,}240 \\
        \bottomrule
    \end{tabular}
\end{table}

At the simulated free-flow speed of 72.4 km/h, the six levels span per-approach flow rates from 724 veh/h at LOS A to 7,240 veh/h at traffic level-of-service F. Table \ref{tab:traffic_scenarios} summarizes all six evaluated density scenarios. Each density level is evaluated under both LOS and NLOS channel configurations, yielding 12 simulation scenarios per cryptographic algorithm and 24 total scenarios. The selection of these six levels is motivated by the need to characterize PQC sidelink performance across a representative and complete range of operating conditions. Traffic level-of-service A and B represent free-flow conditions with sparse vehicular populations and minimal sidelink resource usage, providing the most favorable operating environment for Falcon-512 deployment. Level-of-service C and D represent moderate urban arterial conditions where elevated channel load interacts with Falcon-512's increased subchannel utilization. As a result, these conditions are particularly important for determining the traffic densities at which PQC-related performance degradation becomes observable. Level-of-service E represents at-capacity conditions with significant queuing and high concurrent transmitter density, while level-of-service F represents severely oversaturated conditions that stress-test the sidelink resource pool under worst-case conditions. Evaluating across this full range enables identification of the density threshold at which Falcon-512's payload overhead begins to degrade sidelink performance below safety-critical requirements.

\subsection{Channel Propagation Model}
\label{sec:channel_model}

The radio propagation environment significantly affects sidelink reception performance. Two channel configurations are employed to represent distinct propagation conditions: a LOS model and an NLOS model, both derived from the ITU-R M.2135-1 evaluation guidelines \cite{itur_m2135}.

The LOS configuration uses the Analytical path loss model, which provides a dual-slope distance-dependent attenuation. The path loss $\text{PL}_\text{LOS}$ (in dB) is computed as:
\begin{equation}
    \text{PL}_\text{LOS}(d) \!=\! \begin{cases}
    22 \log_{10}(d) + 28 + 20\log_{10}(f_c), & d < d_\text{BP} \\[4pt]
    40\log_{10}(d) + 7.8 - 18\log_{10}(h_\text{TX}) & \\
    \quad - 18\log_{10}(h_\text{RX}) + 2\log_{10}(f_c), & d \geq d_\text{BP}
    \end{cases}
    \label{eq:pl_los}
\end{equation}
Where $d$ is the transmitter-receiver distance (m), $f_c$ is the carrier frequency in GHz, $h_\text{TX}$ and $h_\text{RX}$ are the effective antenna heights (m), and $d_\text{BP}$ is the breakpoint distance. $d_\text{BP}$ is calculated using \ref{eq:breakpoint}.
\begin{equation}
    d_\text{BP} = \frac{4 \, h_\text{TX} \, h_\text{RX} \, f_c}{c}
    \label{eq:breakpoint}
\end{equation}
With $c = 3 \times 10^8$ m/s, the path loss is lower bounded by the free-space path loss, calculated using \ref{eq:pl_free}:
\begin{equation}
    \text{PL}_\text{free}(d) = 20\log_{10}(d) + 46.4 + 20\log_{10}\!\left(\frac{f_c}{5}\right)
    \label{eq:pl_free}
\end{equation}

The NLOS configuration uses the ITU Urban Microcell (UMi) model \cite{itur_m2135}, which applies higher attenuation to account for building obstruction and multipath scattering. The NLOS path loss is:
\begin{equation}
    \text{PL}_\text{NLOS}(d) = 36.7\log_{10}(d) + 22.7 + 26\log_{10}(f_c)
    \label{eq:pl_nlos}
\end{equation}
valid for $10\text{ m} < d < 5000\text{ m}$. This model produces substantially higher attenuation than the LOS model at the same distance, reflecting the signal degradation caused by urban obstructions at intersection environments where buildings obscure cross-street visibility.

Both channel configurations employ Nakagami-$m$ fading to model small-scale multipath effects. The received signal power envelope follows a Gamma distribution with shape parameter $m$ and scale parameter $\Omega/m$, where $\Omega$ is the average received power. The probability density function of the instantaneous received power $P$ is:
\begin{equation}
    f(P) = \frac{m^m \, P^{m-1}}{\Gamma(m) \, \Omega^m} \exp\!\left(-\frac{m \, P}{\Omega}\right)
    \label{eq:nakagami}
\end{equation}
where $\Gamma(\cdot)$ is the Gamma function. A shape parameter of $m = 1$ is used, which reduces to Rayleigh fading and models a rich scattering environment. Log-normal shadowing with spatial correlation is also enabled.


\subsection{Key Performance Indicators}
\label{sec:kpis}

The following KPIs are used to evaluate the impact of PQC on sidelink performance. These metrics are selected to capture both communication reliability and safety-critical application requirements, as V2X safety applications mandate that BSMs be received reliably within strict latency bounds \cite{sae_j29451}.

\hfill\break%
\noindent\textbf{Packet Delivery Ratio (PDR):}~PDR is computed using the 5G Automotive Association (5GAA) P-190033 sliding-window methodology \cite{5gaa_p190033}. For each transmitter-receiver pair, a centered 5-second evaluation window $[t-2.5,\;t+2.5]$ s is applied to each successfully received packet at time $t$. PDR is then defined as:
    \begin{equation}
        \text{PDR} = \frac{N_{\text{received}}}{N_{\text{transmitted}}} \times 100\%
        \label{eq:pdr}
    \end{equation}
where $N_{\text{received}}$ is the number of successfully received packets and $N_{\text{transmitted}}$ is the total number of packets transmitted by the sender within the same evaluation window. PDR directly quantifies communication reliability. A mean PDR of 90\% or above is the accepted threshold for safety-critical vehicular applications \cite{twardokus2025chasm}.

\hfill\break%
\noindent\textbf{End-to-End Latency:}~End-to-end latency is defined as:
\begin{equation}
    \Delta t_\text{E2E} = T_\text{recv} - T_\text{send}
    \label{eq:latency_def}
\end{equation}
where $T_\text{send}$ is the timestamp recorded at the sender just before the signed SPDU is submitted to the immediate lower layer, i.e., the Packet Data Convergence Protocol (PDCP) layer. $T_\text{recv}$ is the timestamp recorded at the receiver's application layer immediately upon SPDU arrival, before signature verification begins. This formulation captures MAC scheduling delay, the time the SPDU waits for its next SPS grant slot within the 100 ms RRI, physical-layer transmission time, and radio propagation delay. Signature generation at the sender and signature verification at the receiver occur outside the measured interval and are therefore not reflected in $\Delta t_\text{E2E}$, allowing communication-layer delay to be evaluated independently of cryptographic processing overhead. SAE J2945/1 \cite{sae_j29451} requires BSM delivery within 100 ms, making this a hard upper bound for V2X system design.


\hfill\break%
\noindent\textbf{Subchannel Resource Utilization:}~The number of subchannels consumed per transmission for each algorithm at a given MCS. As shown in Table \ref{tab:tbs_feasibility}, PQC algorithms require significantly more subchannels than ECDSA, reducing the spatial multiplexing capacity of the sidelink and increasing contention among concurrent transmitters.

\section{Evaluation Outcomes}
\label{sec:results}

This section presents the simulation results for all 24 scenarios across six traffic level-of-service and two channel configurations. PDR is evaluated using the 5GAA P-190033 sliding-window method, with a centered 5-second evaluation window. End-to-end latency statistics are derived from per-packet application-layer reception logs. Table~\ref{tab:perf_all} summarizes end-to-end latency for all scenarios.

\begin{table}[!t]
\caption{End-to-End Latency Across All 24 Scenarios}
\label{tab:perf_all}
\centering
\scriptsize
\begin{tabular}{llcrrrr}
\toprule
    \textbf{Channel} & \textbf{Traffic level-of-service} & \textbf{Algorithm} & \multicolumn{4}{c}{\textbf{$\Delta t_\text{E2E}$}} \\
    & & & \textbf{Mean (ms)} & \textbf{Median (ms)} & \textbf{P95 (ms)} & \textbf{Max (ms)} \\
\midrule
\multirow{12}{*}{LOS}
 & \multirow{2}{*}{A}  & ECDSA P-256  & 51.71 & 52 & 97.0 & 101 \\
 &                       & Falcon-512   & 52.63 & 53 & 97.0 & 198 \\[3pt]
 & \multirow{2}{*}{B}  & ECDSA P-256  & 51.96 & 52 & 97.0 & 101 \\
 &                       & Falcon-512   & 52.03 & 52 & 97.0 & 198 \\[3pt]
 & \multirow{2}{*}{C}  & ECDSA P-256  & 51.81 & 52 & 96.0 & 101 \\
 &                       & Falcon-512   & 51.79 & 52 & 97.0 & 199 \\[3pt]
 & \multirow{2}{*}{D}  & ECDSA P-256  & 51.54 & 51 & 97.0 & 101 \\
 &                       & Falcon-512   & 51.87 & 51 & 97.0 & 197 \\[3pt]
 & \multirow{2}{*}{E}  & ECDSA P-256  & 51.36 & 51 & 97.0 & 101 \\
 &                       & Falcon-512   & 51.79 & 52 & 97.0 & 199 \\[3pt]
 & \multirow{2}{*}{F}  & ECDSA P-256  & 51.87 & 52 & 97.0 & 101 \\
 &                       & Falcon-512   & 51.88 & 52 & 97.0 & 193 \\
\midrule
\multirow{12}{*}{NLOS}
 & \multirow{2}{*}{A}  & ECDSA P-256  & 50.93 & 52 & 97.0 & 101 \\
 &                       & Falcon-512   & 50.97 & 49 & 98.0 & 197 \\[3pt]
 & \multirow{2}{*}{B}  & ECDSA P-256  & 51.76 & 52 & 98.0 & 101 \\
 &                       & Falcon-512   & 52.15 & 51 & 97.0 & 196 \\[3pt]
 & \multirow{2}{*}{C}  & ECDSA P-256  & 51.95 & 53 & 97.0 & 101 \\
 &                       & Falcon-512   & 52.13 & 52 & 97.0 & 197 \\[3pt]
 & \multirow{2}{*}{D}  & ECDSA P-256  & 49.76 & 50 & 96.0 & 101 \\
 &                       & Falcon-512   & 52.30 & 52 & 97.0 & 194 \\[3pt]
 & \multirow{2}{*}{E}  & ECDSA P-256  & 52.11 & 52 & 96.0 & 101 \\
 &                       & Falcon-512   & 52.66 & 53 & 97.0 & 199 \\[3pt]
 & \multirow{2}{*}{F}  & ECDSA P-256  & 52.14 & 52 & 97.0 & 101 \\
 &                       & Falcon-512   & 52.30 & 52 & 97.0 & 194 \\
\bottomrule
\end{tabular}

\vspace{1mm}
    \footnotesize
    Note: PDR: 5GAA P-190033 sliding window approach with a 5s window \cite{5gaa_p190033}. P95: 95th percentile latency.
\end{table}

\subsection{Packet Delivery Ratio}
\label{sec:pdr_results}

\begin{figure}[t]
    \centering
    \includegraphics[width=\textwidth]{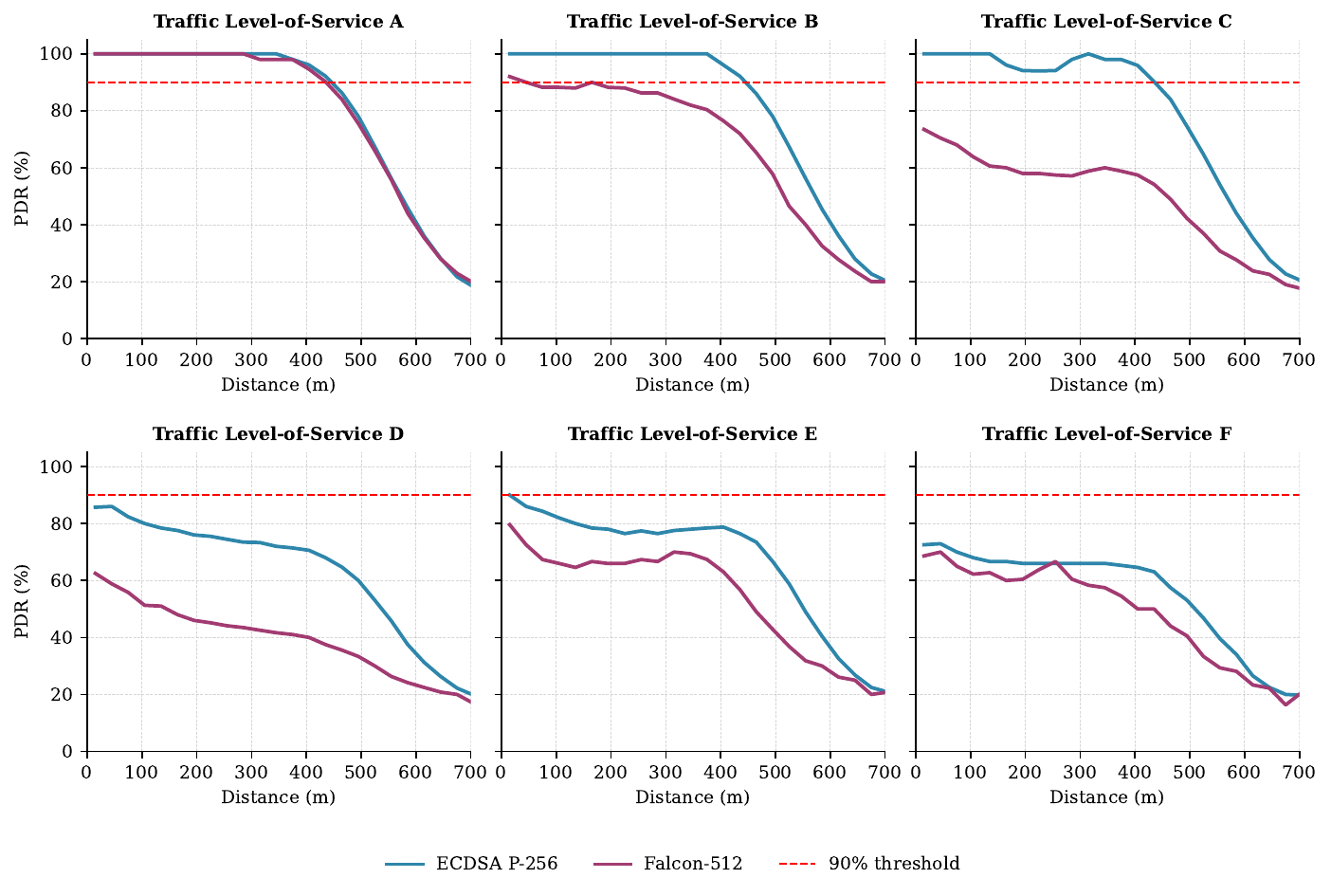}
    \caption{PDR vs.\ sender-RSU distance for ECDSA P-256 and Falcon-512 under LOS channel conditions across all six traffic levels-of-service. The 90\% SAE J2945/1 safety threshold is shown as a red dashed line.}
    \label{fig:pdr_vs_distance_los}
\end{figure}

Figure~\ref{fig:pdr_vs_distance_los} shows PDR as a function of sender-RSU distance for both algorithms across all six traffic levels under LOS propagation. At traffic level-of-service A, the two curves are nearly coincident, and both algorithms deliver above 90\% PDR within around 450m, and both decline at longer distances as path loss attenuates the received signal. The almost identical curves confirm that the sidelink resource pool is lightly loaded enough to absorb Falcon's larger subchannel usage without significant contention. The picture changes at the traffic level-of-service B. The PDR of Falcon is lower than the 90\% boundary even at a close distance of less than 100 m, while ECDSA sustains above 90\% PDR out to approximately 450 m. The gap is most noticeable at traffic level-of-service C. ECDSA maintains 90\% PDR to nearly 450 m, like the previous two scenarios, while Falcon's curve starts near 73\% even at distances below 50 m and never reaches the threshold at any distance. The loss at short range cannot be attributed to propagation, as at these distances the signal is well above the decoding threshold. It is caused entirely by SCI-level interference from competing Falcon reservations occupying 40\% of the subchannel pool per transmission. From traffic level-of-service D onward, neither algorithm meets the 90\% safety threshold at any distance. ECDSA starts at approximately 85\% at short range and declines gradually; Falcon starts near 62\% and follows a steeper descent. The two curves converge near 20\% at distances beyond 500 m for all high-traffic scenarios, reflecting the propagation limit that both algorithms share regardless of their spectral occupancy.

Under NLOS propagation, the PDR vs.\ distance profile differs fundamentally from the LOS case. Under LOS, both algorithms deliver above 90\% PDR at short distances and degrade with range. Under NLOS, neither ECDSA nor Falcon-512 meets the 90\% safety threshold at any communication distance, including ranges as short as 100 m. The ITU Urban Microcell NLOS path loss model produces severe signal attenuation that reduces link reliability throughout the intersection geometry, regardless of the algorithm or traffic level. The consequence is that the relative gap between ECDSA and Falcon-512 narrows substantially under NLOS. The PHY-layer loss attribution confirms the reason. The interference is negligible for both algorithms across all NLOS traffic levels, meaning resource allocation plays no meaningful role. When propagation attenuation governs reception probability on sidelink channel, the number of subchannels a transmitter occupies has little bearing on the outcome. Both algorithms fail for the same reason, at the same distances, and to a similar degree. The broader implication is that in NLOS environments, improving PDR requires infrastructure design: RSU placement, antenna positioning, or repeater deployment, rather than algorithm selection. Switching between ECDSA and Falcon-512 does not materially change the delivery outcome when propagation is the binding constraint.

\subsection{End-to-End Latency}
\label{sec:latency_results}

Table~\ref{tab:perf_all} reports latency statistics for all 24 scenarios. Mean end-to-end latency ranges from 49.76 to 52.66 ms, and median latency ranges from 49 to 53 ms, across all scenarios and both algorithms. The near-coincidence of mean and median in every scenario confirms that the latency distribution is tightly concentrated: the vast majority of packets are delivered well within the SAE J2945/1 100 ms safety deadline, with the 95th percentile latency between 96 and 98 ms across all scenarios. The mean latency difference between ECDSA and Falcon-512 is at most 2.54 ms and below 1 ms in most scenarios. The median difference is at most 4 ms. These results confirm that Falcon-512's computational overhead of signature generation and verification via the liboqs reference library contributes negligibly to end-to-end delay. The dominant latency component is MAC-layer scheduling delay, which is governed by the 100 ms SPS RRI and is identical for both algorithms regardless of payload size.



Taken together, the PDR and latency results establish an important distinction. Falcon-512 satisfies the J2945/1 latency requirement for packets that are successfully delivered: mean and median latency are both approximately 52 ms, and 95\% of delivered packets arrive within 97 to 98 ms. The constraint Falcon fails is reliability, not timeliness. For safety-critical V2X applications, both properties are required simultaneously: messages must be delivered frequently enough and quickly enough to be actionable. The PDR results show that Falcon-512 meets the reliability threshold only at traffic level-of-service A under LOS conditions, establishing spectrum efficiency, not computational capacity, as the primary barrier to PQC adoption in C-V2X sidelink.

\section{Discussion}
\label{sec:discussion}
This section interprets the simulation results in the context of PQC deployment for C-V2X PC5 Mode 4 sidelink. Two questions motivate the analysis. The first aspect of the analysis examines which NIST PQC schemes can be accommodated within the transport block constraints of the SAE J3161 profile. The second examines whether the physically feasible schemes can satisfy the 90\% PDR safety requirement of SAE J2945/1 across realistic traffic conditions. The results show that these two criteria are not equivalent. Falcon-512 satisfies the first but fails the second at all traffic levels above level-of-service A under LOS propagation. The discussion proceeds from feasibility assessment to level-of-service thresholds and deployment implications to trade-offs and mitigation strategies and concludes with the limitations of the study.

\subsection{Feasibility Assessment}

The TBS analysis in this study confirms that Dilithium-2 and SPHINCS+ are physically incompatible with the current SAE J3161 profile. No combination of MCS and subchannel allocation within the J3161-permitted range can accommodate their SPDU sizes, and accommodating them would require protocol changes outside the scope of this study. Falcon-512 is the only NIST PQC scheme that fits within the current SAE J3161 profile. Its certificate-bearing SPDU of 1,735 bytes requires 7 subchannels at MCS 11, and its digest-bearing SPDU of 741 bytes requires only 4 subchannels. In Mode 4 operation, the SPS grant stabilizes at 4 subchannels, as the digest-bearing SPDU is the dominant packet type. However, fitting within the TBS is a necessary condition but not a sufficient one for safe deployment. Under LOS conditions, ECDSA meets the 90\% PDR safety threshold at traffic level-of-service A, B, and C, achieving 94.89\%, 95.40\%, and 95.02\%, respectively. Falcon meets the threshold only at traffic level-of-service A, where it achieves 92.76\%. At traffic level-of-service B, ECDSA achieves 95.40\% while Falcon drops to 79.86\%. This early failure at a moderately free traffic level, before channel saturation sets in, confirms that the performance degradation is not caused by high aggregate load but by Falcon's 4-subchannel grant occupying 40\% of the resource pool on each transmission. Physical transport block compatibility alone does not make a PQC scheme deployable.

\subsection{Level-of-Service Thresholds and Deployment Implications}

Figure~\ref{fig:pdr_vs_distance_los} maps the practical deployment boundary for each algorithm under LOS propagation. ECDSA meets the 90\% PDR safety threshold at traffic level-of-service A, B, and C, sustaining reliable delivery out to approximately 450 m at each of these traffic levels. Falcon-512 meets the threshold only at traffic level-of-service A, where both algorithms produce nearly coincident curves and the resource pool is lightly enough loaded to absorb Falcon's larger subchannel usage without significant contention. From traffic level-of-service B onward, Falcon's PDR falls below the 90\% threshold even at ranges below 100 m, a distance at which propagation loss is negligible and the received signal strength is well above the decoding threshold. This failure at short range confirms that the degradation is not propagation-driven and cannot be recovered by improving channel conditions or reducing communication distance. At traffic level-of-service D and above, neither algorithm meets the threshold at any distance. The two curves converge at high traffic levels, reaching similar low values at traffic level-of-service F, where full channel saturation eliminates the relative benefit of ECDSA's more compact subchannel usage. PDR analysis establishes that under NLOS propagation, neither algorithm meets the 90\% PDR safety threshold at any communication distance. The key deployment implication is that NLOS failures are symmetric: ECDSA performs no better than Falcon-512 under NLOS conditions. This symmetry means NLOS propagation cannot be cited as a specific objection to PQC migration. Any V2X deployment that fails the reliability requirement under NLOS does so regardless of whether it uses ECDSA or Falcon-512. The failure is architectural, not cryptographic. End-to-end latency does not present a deployment barrier for Falcon-512. Mean latency ranges from 49 to 52 ms for both algorithms across all 24 scenarios, and the 95th percentile latency stays within 97 to 98 ms, well under the SAE J2945/1 100 ms deadline. Falcon tail latency can reach 193 to 199 ms in rare cases, reflecting occasional double-cycle queuing when a large transport block misses its scheduled grant slot. These events represent fewer than 1\% of delivered packets and do not affect the 95th percentile. The binding constraint for Falcon-512 is PDR, not timeliness.

The performance of both algorithms under NLOS propagation clarifies the envelope within which Falcon-512's algorithm-specific penalty is relevant. Where LOS coverage is available, the penalty is real and traffic-dependent. Falcon meets the 90\% PDR threshold at traffic level-of-service A and fails above it, while ECDSA sustains reliability to traffic level-of-service C. Where NLOS conditions dominate, algorithm selection becomes irrelevant because neither scheme can meet the requirement regardless of traffic load. Falcon-512's deployability is therefore bounded by LOS coverage availability, not by any inherent property that could be addressed through algorithm optimization. The practical consequence for deployment planning is that coverage design must precede algorithm selection in NLOS environments. RSU placement, antenna orientation, and relay integration are the binding design variables for V2X intersection deployments under NLOS. Achieving reliable sidelink communication under NLOS is a prerequisite for any authentication scheme, PQC or classical, to meet SAE J2945/1 requirements. The question of which scheme to deploy is secondary to whether sufficient LOS coverage exists within the communication zone.

\subsection{Trade-offs and Mitigation Strategies}

The central trade-off is between quantum resistance and sidelink resource efficiency. 
Falcon-512 provides long-term protection against quantum attacks, but its SPS grant occupies 4 subchannels per transmission, twice the resource usage of ECDSA's 2-subchannel grant. Certificate-bearing BSMs, which occur once every five transmissions under the J2945/1 policy, require 7 subchannels and are fragmented by the RLC layer across two transmission periods, incurring elevated loss probability on each certificate exchange. This directly raises persistent collision probability under Mode 4 SPS scheduling, especially at high traffic levels. One path forward is through protocol evolution. SAE J3161 currently limits PSSCH MCS to indices 5, 6, 7, and 11. Extending this range to higher MCS indices would allow the same SPDU to fit into fewer subchannels, reducing Falcon's resource consumption. Validating higher-order modulations, such as 64-QAM and 128-QAM,  under vehicular channel conditions would be a prerequisite before including them in the standard. Packet fragmentation across multiple subframes, as proposed in~\cite{twardokus2025chasm}, is another option, but it adds reassembly complexity and increases sensitivity to individual packet loss events. A hybrid certificate design may offer practical near-term flexibility. Vehicles could carry both ECDSA and Falcon credentials and select the active algorithm based on observed channel conditions. Falcon would be used at low traffic levels where its PDR impact remains within acceptable bounds, and ECDSA would be used when channel load is high. Implementing this approach would require changes to IEEE 1609.2 and SAE J3161 to define switching conditions and to support trust verification under both algorithms simultaneously. Preliminary studies~\cite{bazzi2021sidelink} suggest NR-V2X could accommodate Falcon-512 SPDUs with minimal PDR degradation up to traffic level-of-service E. However, NR-V2X deployment faces regulatory and backwards-compatibility challenges, as most current V2X infrastructure is LTE-based.

\section{Conclusion}
\label{sec:conclusion}

This study evaluated the feasibility of deploying NIST-standardized PQC-DSAs on C-V2X PC5 Mode 4 sidelink communication for safety-critical vehicular applications. A transport block feasibility analysis first determined which algorithms are physically compatible with the SAE J3161 deployment profile. The feasibility analysis showed that Dilithium-2 and SPHINCS+ cannot be accommodated within the current SAE J3161 profile without protocol modifications. Dilithium-2's digest-bearing SPDU of 2,495 bytes exceeds the 2,481-byte TBS ceiling at MCS 11 with all 10 subchannels, and SPHINCS+ at 7,856 bytes is infeasible by a much larger margin. Falcon-512 is the only NIST PQC digital signature scheme compatible with the current SAE J3161 profile: its certificate-bearing SPDU of 1,735 bytes requires 7 subchannels at MCS 11, while in practice the SPS grant settles at a steady state of 4 subchannels, with certificate-bearing transmissions fragmented by the RLC layer across two transmission periods. A co-simulation environment evaluated the performance of Falcon-512 and the baseline ECDSA P-256 across 24 scenarios spanning six traffic level-of-service and two propagation conditions. Falcon-512 satisfies the SAE J2945/1 latency requirement for successfully delivered packets, with mean and median latency near 52 ms and 95th percentile latency within 97 to 98 ms across all 24 scenarios. However, it fails the 90\% PDR reliability requirement at all traffic levels above traffic level-of-service A under LOS propagation, where ECDSA meets the threshold through traffic level-of-service C. Under NLOS propagation, neither algorithm meets the 90\% PDR threshold at any communication distance, including at ranges below 100 m. Severe path loss from the ITU Urban Microcell model dominates packet loss regardless of subchannel allocation or traffic level, making propagation coverage rather than algorithm selection the binding constraint under NLOS.

Several aspects of this study limit the generalizability of the results. The evaluation uses a single four-leg signalized intersection scenario. This geometry produces different collision rates than other intersection geometries or a highway corridor. Liboqs is used for PQC implementation, which is not optimized for embedded automotive hardware. The OpenCV2X simulator does not implement the SPS+One-Shot transmission mechanism required by SAE J3161 Section 7.3.3. This mechanism reduces repetitive scheduling collisions caused by half-duplex operation. Additionally, the OpenCV2X simulator does not implement the subchannel constraint specified in SAE J3161 Section~8.6 (REQ-J3161-SPS\_004), which prohibits allocations occupying more than 50\% but less than 100\% of available subchannels. Under strict J3161 compliance, the 7-subchannel allocation required for Falcon-512 certificate-bearing SPDUs would be rounded up to a full 10-subchannel allocation. The simulation results reflect a 7-subchannel operating point for certificate transmissions that a fully compliant implementation would not use. Future work should examine whether these constraints can be relaxed through protocol and infrastructure changes. Evaluating Falcon-512 on 3GPP Release 16/17 NR-V2X would determine whether higher data rates expand the viable traffic range beyond traffic level-of-service A. Hybrid certificate schemes, in which vehicles switch between Falcon and ECDSA based on observed channel congestion, could provide quantum resistance without sacrificing reliability under heavy load but would require changes to IEEE 1609.2 and SAE J3161 to define switching conditions. Extending the SAE J3161 MCS range to higher indices would reduce Falcon's subchannel requirement and mitigate its resource contention penalty. Multi-intersection corridor simulations and hardware profiling on automotive-grade embedded platforms would provide a more complete picture of deployment feasibility across diverse operating environments.


\section*{FUNDING}
This work is supported by the National Center for Transportation Cybersecurity and Resiliency (TraCR) (a U.S. Department of Transportation National University Transportation Center) headquartered at Clemson University, Clemson, South Carolina, USA.

\section{Acknowledgments}
Any opinions, findings, conclusions, and recommendations expressed in this material are those of the author(s) and do not necessarily reflect the views of funding agencies, and the U.S. Government assumes no liability for the contents or use thereof.
Also, the authors used Anthropic Claude and OpenAI ChatGPT for editorial purposes

\section*{AUTHOR CONTRIBUTIONS}
The authors confirm contribution to the paper as follows: 
study conception and design: Akid Abrar, Sagar Dasgupta, and Mizanur Rahman;
data collection: Akid Abrar, Minhaj Uddin Ahmad;
analysis and interpretation of results: Akid Abrar, Sagar Dasgupta, Abdullah Al Mamun, Minhaj Uddin Ahmad, Mizanur Rahman; Mashrur Chowdhury and Ahmad Alsharif; 
draft manuscript preparation: Akid Abrar, Sagar Dasgupta, Abdullah Al Mamun, Minhaj Uddin Ahmad, and Mizanur Rahman;
funding acquisition, review, and editing: Sagar Dasgupta, Mizanur Rahman, Mashrur Chowdhury, and Ahmad Alsharif.
All authors reviewed the results and approved the final version of the manuscript.

\section*{DECLARATION OF CONFLICTING INTERESTS}
The authors declared no potential conflicts of interest with respect to the research, authorship, and/or publication of this article.

\printbibliography[title={REFERENCES}]
\end{document}